\documentclass[12pt]{article}
\usepackage{graphicx}
\usepackage{amsfonts}
\usepackage{amssymb,amsmath,bm}
\usepackage{latexsym}
\usepackage{color}
\usepackage{cite}
\input{colordvi.tex}
\usepackage[utf8]{inputenc}
\begin{document}
\setcounter{footnote}{0}

\begin{titlepage}

\begin{center}
\vskip .5in

{\Large \bf
Clustering of primordial black holes
with non-Gaussian initial fluctuations
}
\vskip .45in

{
Teruaki Suyama$^{a}$,
and Shuichiro Yokoyama$^{b,c}$
}


{\em
$^a$
   Department of Physics, Tokyo Institute of Technology, 2-12-1 Ookayama,
   Meguro-ku, Tokyo 152-8551, Japan
   }\\

{\em
$^b$
  Kobayashi Maskawa Institute,
Nagoya University,
Aichi 464-8602, Japan
}\\

{\em
$^c$
Kavli IPMU (WPI), UTIAS, The University of Tokyo,
Kashiwa, Chiba 277-8583, Japan
}\\

\end{center}

\vskip .4in

\begin{abstract}%
We formulate the two-point correlation function of primordial black holes (PBHs) at their formation time,
based on the functional integration approach which has often been used in the context of
halo clustering. We find that PBH clustering on super-Hubble scales could never be induced
in the case where the initial primordial fluctuations are Gaussian, while it can be enhanced by
the so-called local-type trispectrum (four-point correlation function) of the primordial curvature perturbations. 
\end{abstract}
\end{titlepage}

\section{Introduction}
\label{sec:intro}

Thanks to the recent detections of gravitational waves from binary black holes by LIGO/VIRGO collaboration,
primordial black holes (PBHs) are attracting attention as a candidate for such binary black hole systems.
In Ref.~\cite{Sasaki:2016jop} (see also Ref.~\cite{Ali-Haimoud:2017rtz}), we estimated the merger rate of PBH binary systems
and found that if the primordial black holes account for $0.1\% \sim 1.0\%$ of the dark matter in the Universe,
the PBH scenario can be consistent with the first LIGO gravitational wave (GW) event, GW150914.
In the analysis in Ref. \cite{Sasaki:2016jop}, we have assumed that the distribution of PBHs is spatially uniform.
As discussed in Ref.~\cite{Raidal:2017mfl} (see also~\cite{Ballesteros:2018swv,Bringmann:2018mxj,Ding:2019tjk}),\footnote{ 
Clustering of primordial black holes has also been discussed in 
Refs.~\cite{Dokuchaev:2008hz,Belotsky:2018wph}.
These papers have investigated the observational effects of PBH clustering, not only on the binary formation but also on the formation of the supermassive BHs.} if PBHs spatially clustered at the formation, which can be characterized by the
two-point correlation function, it would affect the probability of PBH binary formation and the estimation
of the merger rate of PBH binaries. 

There are several works about the initial clustering of PBHs.\footnote{There have been several works about late-time clustering of PBHs, e.g. in dark matter halos (see, e.g., Refs.~\cite{Clesse:2016vqa,Garcia-Bellido:2017xvr}).}
As a pioneer work, Ref.~\cite{Meszaros:1975ef} (and also Ref.~\cite{Carr:1975qj}) discussed galaxy formation due to the spatial fluctuations in PBH number density.
More detailed discussion was given in Ref.~\cite{Chisholm:2005vm}.
The PBH two-point correlation function and the power spectrum
were  estimated in Ref.~\cite{Chisholm:2005vm}
by making use of the peak formalism developed in the context of the clustering of galaxies/halos.
As a result, the power spectrum of PBH distribution on large scales should be dominated by the Poisson noise and it behaves
as matter isocurvature perturbations.
Based on such a PBH isocurvature due to the Poisson noise,
Ref.~\cite{Afshordi:2003zb} put a constraint on the abundance of PBHs
from Ly-$\alpha$ forest observations, and also 
Refs.~\cite{Gong:2017sie,Gong:2018sos} estimated an expected constraint on it by making use of future 21cm observations.
Recently, Ref.~\cite{Ali-Haimoud:2018dau} studied the initial clustering of PBHs
in more detail and found that even on much smaller scales PBHs should not be clustered
beyond Poisson. Furthermore, Ref.~\cite{Desjacques:2018wuu}
investigated the dependence of the clustering feature
on the shape of the primordial curvature power spectrum.
While Refs.~\cite{Chisholm:2005vm, Ali-Haimoud:2018dau Desjacques:2018wuu}
assumed Gaussian initial fluctuations,
Refs.~\cite{Tada:2015noa, Young:2015kda} investigated the impact of primordial local-type non-Gaussianity on the super-Hubble density fluctuations of PBHs,
based on the peak-background split picture.
Although they found that local-type non-Gaussianity
could induce the super-Hubble correlations of the PBH density fluctuations,
they focused on a specific type of non-Gaussianity
and did not obtain a formula for
a PBH two-point correlation function which is
applicable for more general types of non-Gaussianity.
Recently, Ref.~\cite{Franciolini:2018vbk} also investigated 
the effect of primordial non-Gaussianities
not only on the abundance but also on the clustering property of PBHs.
In Ref.~\cite{Franciolini:2018vbk}, the threshold of PBH formation is supposed to be
given in terms of primordial curvature perturbations.
Their result indicates that PBHs are clustered beyond Poisson on 
super-Hubble scales
even for Gaussian primordial curvature perturbations,
which would be apparently inconsistent with the results obtained in previous works.

In this paper we estimate the two-point correlation function of PBHs by making use of a functional integration approach.
This method is powerful in studying correlation functions
of biased objects since it allows us to systematically include the effect of non-Gaussian 
properties of the underling density fluctuations\cite{Matarrese:1986et,Matarrese:2008nc,Gong:2011gx}.
Actually, this approach has also been
used in Ref.~\cite{Franciolini:2018vbk}.
In the radiation-dominated era,
PBHs are actually formed soon after horizon reentry
if the amplitude of primordial fluctuations is greater than a certain threshold.
The primordial fluctuation often used to study a criterion for PBH formation
is the density contrast in a comoving slice.
This quantity represents a local three-curvature and is in good accordance with
the physical argument that PBH formation should be determined
by local dynamics (see, {\it e.g.}, Ref. \cite{Young:2014ana}).
Thus, contrary to Ref.~\cite{Franciolini:2018vbk}, 
we employ the density contrast in a comoving slice
as a critical quantity for PBH formation.

This paper is organized as follows.
In the next section, based on the functional integration approach (or path integral method: see, e.g., Ref.~\cite{Matarrese:1986et}), 
we formulate a two-point correlation function for PBHs
which is applicable to non-Gaussian primordial perturbations.
In Sect.~\ref{sec:local} we investigate
the possibility that PBHs are clustered on large scales due to primordial non-Gaussianities.
In order to discuss how PBHs are clustered,
we simply assume scale-independent local-type non-Gaussianity of the primordial
curvature perturbations and show the relation between the PBH two-point correlation on large scales and the non-linearity parameter.
Section~\ref{sec:con} is devoted to the conclusion.

\section{Formulation}
\label{sec:form}

Since 
during the radiation-dominated era PBHs are formed in the overdense (or positive spatial curvature) region with the Hubble horizon size, the criterion for PBH formation is supposed to be locally determined independently of super-Hubble scale fluctuations.
Thus we introduce {\it local} smoothed primordial fluctuations $\theta_{\rm local}({\bm x})$ as 
\begin{equation}
\theta_{\rm local} ({\bm x}) := \int d^3 y\, W_{\rm local} ({\bm x} - {\bm y}) ~\theta ({\bm y}),
\label{eq:local}
\end{equation}
where the window function $W_{\rm local}({\bm x})$ is a smoothing 
function with scale $R$ that also
removes wavelength modes longer
than the scale $R$.
For PBH formation, the scale $R$ is roughly matched with the Hubble horizon size at the formation.
In this section, to maintain generality
we do not specify a particular gauge for defining the primordial fluctuations $\theta ({\bm x})$, 
and
assume the criterion for PBH formation is given by\footnote{Strictly speaking, the threshold depends on the perturbation profile (see, e.g., Ref.~\cite{Nakama:2013ica}).
In this paper we ignore this dependence and assume that the threshold is the same for all profiles. 
This assumption does not affect the main result of this paper.}
\begin{equation}
\theta_{\rm local} \geq \theta_{\rm th}.
\end{equation}
In the next section we will choose the density contrast 
on the comoving slice as $\theta_{\rm local}({\bm x})$.

\subsection{Probability of PBH formation}

The probability that a point ${\bm x}$ becomes a PBH can be given by
\begin{equation}
P_1({\bm x}) =  \int [D\theta] P[\theta] \int_{\theta_{\rm th}}^{\infty} d\alpha\, \delta_D (\theta_{\rm local}({\bm x}) - \alpha),
\end{equation}
where $P[\theta]$ is a probability distribution function for the primordial fluctuations, $\theta ({\bm x})$;
also, the probability that two points ${\bm x}_1$ and ${\bm x}_2$ are PBHs can be given by
\begin{equation}
P_2 ({\bm x}_1,\,{\bm x}_2) = \int [D\theta] P[\theta] \int_{\theta_{\rm th}}^{\infty} d\alpha_1\, \delta_D (\theta_{\rm local}({\bm x}_1) - \alpha_1)
\int_{\theta_{\rm th}}^{\infty} d\alpha_2\, \delta_D (\theta_{\rm local}({\bm x}_2) - \alpha_2).
\end{equation}
By using the expression for the one-dimensional Dirac delta function given as
\begin{equation}
\delta_D (x) = \int \frac{d\phi}{2 \pi} e^{i \phi x},
\end{equation}
and the expression for the local smoothed fluctuations given by Eq. (\ref{eq:local}), we have
\begin{equation}
P_1({\bm x}) =  \int [D\theta] P[\theta] \int_{\theta_{\rm th}}^{\infty} d\alpha\,\int^\infty_{-\infty} \frac{d \phi}{2\pi} 
\exp \left[ i \phi \int d^3 y \,W_{\rm local} ({\bm x} - {\bm y}) \,\theta ({\bm y}) - i \phi \, \alpha \right]
\end{equation}
and
\begin{eqnarray}
P_2 ({\bm x}_1,\,{\bm x}_2) &=& \int [D\theta] P[\theta] \int_{\theta_{\rm th}}^{\infty} d\alpha_1\,\int_{\theta_{\rm th}}^{\infty} d\alpha_2\,
\int^\infty_{-\infty} \frac{d \phi_1}{2\pi} \int^\infty_{-\infty} \frac{d \phi_2}{2\pi} \exp \left[ - i \phi_1\,\alpha_1 - i \phi_2 \, \alpha_2 \right] 
\cr\cr
&&\times \exp \left\{ i \int d^3 y \left[ \phi_1 W_{\rm local} ({\bm x}_1 - {\bm y}) + \phi_2 W_{\rm local} ({\bm x}_2 - {\bm y}) \right] \theta ({\bm y}) \right\}\,.
\end{eqnarray}

\subsection{$P_1$ and $P_2$ in terms of the $n$-point correlators of the primordial fluctuations}

Let us introduce a generating function given by
\begin{equation}
Z[J] := \int [D \theta] P[\theta] \exp \left[ i \int d^3 y J({\bm y})  \theta ({\bm y}) \right] = \langle  \exp \left[ i \int d^3 y J({\bm y})  \theta ({\bm y}) \right] \rangle \,,
\end{equation}
with $Z[0] = 1$.
The ``connected" $n$-point correlation functions of $\theta ({\bm x})$ can be given in terms of the generating function as
\begin{equation}
\xi_{\theta (c)} ({\bm x}_1, {\bm x}_2, ..., {\bm x}_n) := \frac{1}{i^n} \frac{ \delta^n \log Z[J]}{\delta J({\bm x}_1) \delta J({\bm x}_2) \cdots \delta J ({\bm x}_n)} \biggr|_{J = 0}\,.
\end{equation}
Inversely, we can obtain the expression for $\log Z[J]$ in terms of the ``connected" $n$-point correlation functions as
\begin{equation}
\log Z[J] = \sum^\infty_{n=1} \frac{i^n}{n!} \int d^3 y_1 d^3 y_2 \cdots d^3 y_n \, \xi_{\theta (c)} ({\bm y}_1,{\bm y}_2, \cdots, {\bm y}_n) \,
J({\bm y}_1) J({\bm y}_2) \cdots J({\bm y}_n) \, . 
\end{equation}
Choosing $J({\bm y}) : = \phi~ W_{\rm local} ({\bm x} - {\bm y})$,
the one-point probability of PBH formation, $P_1$, can be written as
\begin{eqnarray}
P_1 ({\bm x}) &=& \int_{\theta_{\rm th}}^{\infty} d\alpha\,\int^\infty_{-\infty} \frac{d \phi}{2\pi} e^{-i\phi\,\alpha}
\,Z\left[\phi\, W_{\rm local} ({\bm x} - {\bm y})\right] \cr\cr
 &=& \int_{\theta_{\rm th}}^{\infty} d\alpha\,\int^\infty_{-\infty} \frac{d \phi}{2\pi} \exp \left[ - i \phi \, \alpha \right] 
 \exp\left[ \sum^\infty_{n=2} \frac{i^n}{n!} \phi^n \, \xi^{(n)}_{{\rm local} (c)}  \right] \, ,
\end{eqnarray}
where $\xi^{(n)}_{{\rm local} (c)}$ correspond to the moments of $\theta_{\rm local}$:
\begin{equation}
\xi_{{\rm local} (c)}^{(n)}  := \int d^3 y_1 d^3 y_2 \cdots d^3 y_n \, \xi_{\theta (c)} ({\bm y}_1,{\bm y}_2, \cdots, {\bm y}_n)
\prod^n_{r=1} W_{\rm local} ({\bm x} - {\bm y}_r)~.
\end{equation}

In the same way, 
the two-point probability of PBH formation, $P_2$, can be expressed as
\begin{eqnarray}
P_2 ({\bm x}_1,\,{\bm x}_2) &=&  \int_{\theta_{\rm th}}^{\infty} d\alpha_1\,\int_{\theta_{\rm th}}^{\infty} d\alpha_2\,
\int^\infty_{-\infty} \frac{d \phi_1}{2\pi} \int^\infty_{-\infty} \frac{d \phi_2}{2\pi} \exp \left[ - i \phi_1\,\alpha_1 - i \phi_2 \, \alpha_2 \right] \cr\cr
&& \qquad \times \exp \left[ \sum^\infty_{n=2} i^n \sum_{m = 0}^n \frac{\phi_1^m \phi_2^{n-m}}{m! (n-m)!} \xi_{{\rm local}(c),m}^{(n)}\right] \, ,
\end{eqnarray}
where
\begin{eqnarray}
\xi_{{\rm local}(c),m}^{(n)} & := &\xi_{{\rm local}(c)}^{(n)}(
\underbrace{{\bm x}_1, {\bm x}_1, \cdots, {\bm x}_1}_{{\rm total}~m},\underbrace{{\bm x}_2, {\bm x}_2, \cdots {\bm x}_2}_{{\rm total}~n-m}) \cr\cr
& = & \int d^3 y_1 d^3 y_2 \cdots d^3 y_n \, \xi_{\theta (c)} ({\bm y}_1,{\bm y}_2, \cdots, {\bm y}_n) \cr\cr
&& \qquad\qquad \times
\prod^m_{r_1 = 1} W_{\rm local} ({\bm x}_1 - {\bm y_{r_1}})\prod^n_{r_2 = m+1} W_{\rm local}({\bm x}_2 - {\bm y}_{r_2})\,. \nonumber \\
\end{eqnarray}
This can be regarded as a cross correlation between the $m$th and $(n-m)$th moments of the local smoothed primordial fluctuations.

In the above expression for $P_1$, we can perform the integration with respect to $\phi$,
as
\begin{eqnarray}
\int^\infty_{-\infty} \frac{d \phi}{2\pi} e^{- i \phi \, \alpha } 
 \exp\left[ \sum^\infty_{n=2} \frac{i^n}{n!} \phi^n \, \xi^{(n)}_{{\rm local} (c)}  \right] 
 &=& \int^\infty_{-\infty} \frac{d \phi}{2\pi} 
 \exp\left[ \sum^\infty_{n=3} \frac{i^n}{n!} \phi^n \, \xi^{(n)}_{{\rm local} (c)}  \right] e^{ - i \phi \, \alpha - \frac{1}{2} \phi^2 \sigma_{\rm local}^2} \cr\cr
 &=& \int^\infty_{-\infty} \frac{d \phi}{2\pi} 
 \exp\left[ \sum^\infty_{n=3} \frac{i^n}{n!} \frac{ \xi^{(n)}_{{\rm local} (c)}\partial^n}{\partial (-i\alpha)^n}  \right] e^{ - i \phi \, \alpha - \frac{1}{2} \phi^2 \sigma_{\rm local}^2}\,, \nonumber\\
  &=& \exp\left[ \sum^\infty_{n=3} \frac{(-1)^n}{n!} \xi^{(n)}_{{\rm local} (c)} \frac{d^n}{d \alpha^n}  \right] 
  \frac{e^{-\frac{\alpha^2}{2 \sigma_{\rm local}^2}}}{ \sqrt{2 \pi \sigma_{\rm local}^2}}\, , \nonumber\\
\end{eqnarray}
and then the expression for $P_1$ can be reduced to
\begin{eqnarray}
\label{eq:exP1_1}
P_1 =   \frac{1}{\sqrt{2 \pi}}\int^\infty_{\nu} dw \exp\left[ \sum^\infty_{n=3} \frac{(-1)^n}{n!}\frac{\xi^{(n)}_{{\rm local} (c)}}{\sigma_{\rm local}^n}  \frac{d^n}{d w^n}  \right] 
 e^{-\frac{w^2}{2}}
  \, ,
\end{eqnarray}
where $w := \alpha / \sigma_{\rm local}$, $\nu := \theta_{\rm th} / \sigma_{\rm local}$, and $\sigma_{\rm local}:= \left(\xi^{(2)}_{{\rm local}(c)}\right)^{1/2}$.
We can also obtain a reduced form for $P_2$:
\begin{eqnarray}
\label{eq:exP2_1}
P_2 ({\bm x}_1,\,{\bm x}_2) 
&=& \frac{1}{2 \pi} \int_{\nu}^{\infty} dw_1\,\int_{\nu}^{\infty} dw_2 
\exp\left[ \frac{\xi_{{\rm local}(c)}^{(2)}({\bm x}_1, {\bm x}_2)}{ \sigma_{\rm local}^2} \frac{\partial^2}{\partial w_1 \partial w_2}
\right. \cr\cr
&& \qquad\qquad\qquad \left.
+ \sum^\infty_{n = 3} (-1)^n\sum_{m=0}^n \frac{\xi_{{\rm local}(c),m}^{(n)}/\sigma_{\rm local}^n}{m! (n-m)!} \frac{\partial^n}{\partial w_1^m \partial w_2^{n-m}}\right]
e^{- \frac{w_1^2 + w_2^2}{2}} \, .
\end{eqnarray}

\subsection{PBH two-point correlation function }

Let us obtain an approximate formula for the PBH two-point correlation
function,
by performing the integration with respect to $w$ in the above expressions.
To do so,
we employ two approximations: (a) a weak non-Gaussian limit and (b) a high peak limit.

In the weak non-Gaussian limit, we assume that 
$\xi^{(n)}_{{\rm local}(c)}/\sigma_{\rm local}^n \ll 1$ for $n \geq 3 $.
Under this assumption, up to the linear order in $\xi^{(n)}_{{\rm local}(c)}/\sigma_{\rm local}^n$ we have
\begin{eqnarray}
P_1 &=& \frac{1}{\sqrt{2 \pi}} \int^\infty_\nu dw \left[1 + \sum_{n=3}^\infty \frac{(-1)^n}{n!}\frac{\xi^{(n)}_{{\rm local} (c)}}{\sigma_{\rm local}^n}  \frac{d^n}{d w^n} \right]  e^{-\frac{w^2}{2}} + O\left( (\xi^{(n)}_{{\rm local} (c)}/\sigma_{\rm local}^n)^2 \right) ~.
\end{eqnarray}
With the assumptions $\xi^{(n)}_{{\rm local}(c),m}/\sigma_{\rm local}^n \ll 1$ for $n \geq 3 $ and $\xi_{{\rm local}(c)}^{(2)}({\bm x}_1, {\bm x}_2) / \sigma_{\rm local}^2 \ll 1$, we can also obtain
\begin{eqnarray}
P_2 &=& \frac{1}{2 \pi} \int^\infty_\nu dw_1  \int^\infty_\nu dw_2 \left[1 + 
\frac{\xi_{{\rm local}(c)}^{(2)}({\bm x}_1, {\bm x}_2)}{ \sigma_{\rm local}^2} \frac{\partial^2}{\partial w_1 \partial w_2} \right. \cr\cr
&&\qquad\qquad \qquad \qquad \qquad \left. +
\sum^\infty_{n = 3} (-1)^n\sum_{m=0}^n \frac{\xi_{{\rm local}(c),m}^{(n)}/\sigma_{\rm local}^n}{m! (n-m)!} \frac{\partial^n}{\partial w_1^m \partial w_2^{n-m}}\right]
e^{- \frac{w_1^2 + w_2^2}{2}} \cr\cr 
&& \qquad \qquad \qquad \qquad \qquad \qquad \qquad \qquad \qquad \qquad
+\, O\left( (\xi^{(n)}_{{\rm local} (c)}/\sigma_{\rm local}^n)^2 \right) \,.
\end{eqnarray}
Furthermore, by making use of the Hermite polynomials, $H_n(x)$
given as $H_n(x) := (-1)^n e^{x^2} (d/dx)^n e^{-x^2}$, 
we can replace the derivatives in terms of $w$, with the Hermite polynomials.
Then, 
finally, by employing the high peak approximation,
$\nu \gg 1$, which might naturally be valid for the PBH formation,
we can perform the integration approximately in terms of $w$ and obtain approximate expressions for $P_1$ and $P_2$ as
\begin{equation}
\label{eq:exP1_a}
P_1 \approx \frac{e^{-\nu^2/2}}{\sqrt{2 \pi} \nu} \left[1 + \sum_{n=3}^\infty \frac{1}{2^{n/2}n!}\frac{\xi^{(n)}_{{\rm local} (c)}}{\sigma_{\rm local}^n}  H_n\left(\frac{\nu}{\sqrt{2}} \right) \right]~,
\end{equation}
and
\begin{eqnarray}
\label{eq:exP2_a}
P_2 ({\bm x}_1, {\bm x}_2) &\approx&
\frac{e^{-\nu^2}}{2 \pi \nu^2} \left[ 1 + \frac{\xi_{{\rm local}(c)}^{(2)} ({\bm x}_1, {\bm x}_2)}{2 \sigma_{\rm local}^2} H_1 \left( \frac{\nu}{\sqrt{2}} \right)^2 \right. \cr\cr
&& \qquad\qquad \left.
+ \sum_{n=3}^{\infty}\sum_{m=0}^{n} \frac{\xi_{{\rm local}(c),m}^{(n)}/\sigma_{\rm local}^n}{2^{n/2}m! (n-m)!}H_m\left(\frac{\nu}{\sqrt{2}} \right)H_{n-m}\left(\frac{\nu}{\sqrt{2}} \right)
 \right]~.
\end{eqnarray}
Then, by using these expressions for $P_1$ and $P_2$, 
the two-point correlation function of the PBHs can be evaluated as
\begin{eqnarray}
\xi_{\rm PBH} ({\bm x}_1, {\bm x}_2) &:=& \frac{P_2({\bm x}_1,{\bm x}_2) }{P_1^2} - 1 \cr\cr
&\approx& \frac{H_1(\nu/\sqrt{2})^2}{2 \sigma_{\rm local}^2} \xi_{{\rm local}(c)}^{(2)} ({\bm x}_1, {\bm x}_2)
+ \sum_{n=3}^{\infty}\sum_{m=1}^{n-1} \frac{H_m(\nu/\sqrt{2})H_{n-m}(\nu/\sqrt{2})}{2^{n/2}m! (n-m)!\sigma_{\rm local}^n}\xi_{{\rm local}(c),m}^{(n)}\,. \nonumber\\
\end{eqnarray}
Note that, strictly speaking, in order for the above expression to be valid, we should require stronger assumptions given as
\begin{eqnarray}
\frac{\xi^{(n)}_{{\rm local}(c),m}}{\sigma_{\rm local}^n}  \ll  \frac{2^{n/2} m!(n-m)!}{H_m (\nu/\sqrt{2}) H_{n-m}(\nu/\sqrt{2})},
\label{eq:assum_1}
\end{eqnarray}
and 
\begin{equation}
\frac{\xi^{(2)}_{{\rm local}(c)}({\bm x}_1,{\bm x}_2)}{\sigma_{\rm local}^2}  \ll  \frac{2}{H_1 (\nu/\sqrt{2})^{2}}.
\label{eq:assum_2}
\end{equation}

Up to the four-point correlation function of the local smoothed primordial fluctuations,
by using an approximate form of the Hermite polynomials, $H_n (x) \sim  2^n x^n $ for $x \gg 1$,
we have 
\begin{eqnarray}
\xi_{\rm PBH} ({\bm x}_1, {\bm x}_2) &\sim& \frac{\nu^2}{\sigma_{\rm local}^2} \xi_{{\rm local}(c)}^{(2)}({\bm x}_1, {\bm x}_2) \cr\cr
&&
+ \frac{1}{2} \frac{\nu^3}{\sigma_{\rm local}^3} \left( \xi^{(3)}_{{\rm local}(c)} ({\bm x}_1, {\bm x}_1, {\bm x}_2)  +  ({\bm x}_1 \leftrightarrow {\bm x}_2) \right) \cr\cr
&&  + \frac{1}{4} \frac{\nu^4}{\sigma_{\rm local}^4} \xi^{(4)}_{{\rm local}(c)} ({\bm x}_1, {\bm x}_1, {\bm x}_2,{\bm x}_2) \cr\cr
&& 
+ \frac{1}{6} \frac{\nu^4}{\sigma_{\rm local}^4} \left( \xi^{(4)}_{{\rm local}(c)} ({\bm x}_1, {\bm x}_2, {\bm x}_2,{\bm x}_2) +  ({\bm x}_1 \leftrightarrow {\bm x}_2) \right). \nonumber\\
\label{eq:PBHcorr}
\end{eqnarray}
This expression corresponds to the result obtained in Refs.~\cite{Matarrese:2008nc,Gong:2011gx} in the context of the halo bias.
Under the assumptions given by Eqs. (\ref{eq:assum_1}) and (\ref{eq:assum_2}), this equation is valid for $\xi_{\rm PBH}({\bm x}_1, {\bm x}_2) \ll 1$.

\section{PBH clustering with local-type non-Gaussianity}
\label{sec:local}

Before a quantitative discussion for the amplitude of the PBH two-point correlation function, 
let us first provide an intuitive idea of why 
we take into account up to the four-point correlation function of the primordial fluctuations in the above formulation.

Hereafter, as the primordial fluctuations, $\theta ({\bm x})$, in the above formulation,
we use the primordial curvature perturbations on the comoving slice, often denoted by ${\mathcal R}_c({\bm x})$.
In the long-wavelength approximation, 
comoving density fluctuations can be given in terms of ${\mathcal R}_c({\bm x})$ as
~\cite{Harada:2015yda}
\begin{eqnarray}
 \delta ({\bm x}) = -\frac{4 (1 + w)}{3 w + 5}  e^{- 5 {\mathcal R}_c({\bm x})/2} \frac{\triangle}{a^2 H^2} ( e^{{\mathcal R}_c({\bm x})/2}), 
 \label{eq:delta}
\end{eqnarray} 
where $w$, $a$ and $H$ are respectively an equation of state of the Universe, a scale factor and the Hubble parameter.
As can be seen in the above expression, if we use the primordial curvature perturbations on the comoving slice 
as $\theta ({\bm x})$ in the previous section, 
a natural variable for the {\it local} primordial fluctuations $\theta_{\rm local}$ would be the comoving
density fluctuations. In fact, this quantity represents a local three-curvature 
and is in good accordance with a physical argument that
the criterion for PBH formation should be determined by local dynamics (i.e. within the Hubble horizon)
and be free from the addition of super-Hubble modes.
By introducing a {\it local} scale factor $a \, e^{{\mathcal R}_c ({\bm x})} \to a$~\cite{Young:2014ana}, at the linear order,
the above expression can be reduced to
\begin{eqnarray}
\delta ({\bm x}) = -\frac{4}{9} \frac{1}{a^2 H^2}  \triangle {\mathcal R}_c({\bm x})~.
\label{eq:pertDelta}
\end{eqnarray}
Here, we take $w = 1/3$ in the radiation-dominated era.
The two-point correlation function of $\delta$ is given by
\begin{equation}
\langle \delta ({\bm x}) \delta ({\bm y}) \rangle={\left( \frac{4}{9a^2 H^2} \right)}^2
\triangle^2 \langle {\cal R}_c ({\bm x}) {\cal R}_c ({\bm y}) \rangle. \label{dxdy}
\end{equation}
Because of the two Laplacians, $\langle \delta ({\bm x}) \delta ({\bm y}) \rangle$ rapidly approaches zero 
for $| {\bm x}-{\bm y}| \gg {(a H)}^{-1}$ unless ${\cal R}_c$ is extremely red-tilted.
Thus, in general, $\langle \delta ({\bm x}) \delta ({\bm y}) \rangle$ is suppressed on super-Hubble scales.

Due to the locality, at the leading order in $\delta$
the PBH abundance at point ${\bm x}$ would be 
determined by the local variance 
$\langle \delta^2 ({\bm x}) \rangle$, and then
the PBH two-point correlation function is given by 
the its correlation. 
If $\delta$ is Gaussian, the correlation of the local variance is given as
\begin{equation}
\frac{\langle \delta^2 ({\bm x})  \delta^2 ({\bm y}) \rangle}{{\langle \delta^2 ({\bm x}) \rangle}^2}-1
=2 {\langle \delta ({\bm x}) \delta ({\bm y}) \rangle}^2~,
\end{equation}
and it should be suppressed on super-Hubble scales, as can be seen by Eq.~(\ref{dxdy}). 
Thus, PBHs are produced by the same amount in every super-Hubble size region.
In other words, PBHs are not clustered on super-Hubble scales.
If, on the other hand, $\delta$ is non-Gaussian, the correlation of the local variance
may remain on super-Hubble scales.
In order to see this explicitly, let us focus on the simple case where ${\cal R}_c$ consists
of two uncorrelated Gaussian fields $\phi,\chi$ as
\begin{equation}
{\cal R}_c ({\bm x})=(1+\alpha \chi ({\bm x})) \phi ({\bm x}). \label{p-tnl}
\end{equation}
Here it is assumed that $\chi$ has super-Hubble scale correlation
and $\phi$ gives a dominant contribution to PBH formation.
For such a case, the density contrast on the comoving slice can be given as
\begin{equation}
\delta ({\bm x}) = -(1 + \alpha \chi({\bm x}) ) \frac{4}{9} \frac{1}{a^2 H^2} \triangle \phi ({\bm x}).
\end{equation}
For super-Hubble distance $| {\bm x}-{\bm y}| \gg {(a H)}^{-1}$, we obtain
\begin{equation}
\frac{\langle \delta^2 ({\bm x})  \delta^2 ({\bm y}) \rangle}{{\langle \delta^2 ({\bm x}) \rangle}^2}-1
\approx 4\alpha^2 \langle \chi ({\bm x}) \chi ({\bm y}) \rangle+{\cal O}(\alpha^4). \label{scl}
\end{equation}
There are two remarks regarding this result.
First, on super-Hubble scales the correlation of the local variance is directly proportional to the correlation function of $\chi$. 
This result simply reflects our naive intuition that a local quantity can possess correlation over super-Hubble distance only when 
the quantity is sourced by another quantity having correlation over super-Hubble distance.
Secondly, the correlation of the local variance is proportional to a part of the connected part of the four-point function of ${\cal R}_c$. 
More explicitly, the connected part of the four-point function of ${\cal R}_c$ is given by
\begin{equation}
\frac{\langle {\cal R}_c^2  ({\bm x}) {\cal R}_c^2  ({\bm y}) \rangle_c}{{\langle {\cal R}_c^2 ({\bm x}) \rangle}^2}
=4\alpha^2 \left[ \langle \chi ({\bm x}) \chi ({\bm y}) \rangle+
\frac{{\langle {\cal R}_c ({\bm x}) {\cal R}_c ({\bm y}) \rangle}^2}{{\langle {\cal R}_c^2 ({\bm x}) \rangle}^2} \left( \langle \chi^2 ({\bm x}) \rangle
+  \langle \chi ({\bm x}) \chi ({\bm y}) \rangle \right) \right] +{\cal O}(\alpha^4),
\end{equation}
and the right-hand side of Eq.~(\ref{scl}) is obtained by ignoring terms proportional to the two-point correlation function of ${\mathcal R}_c$, $\langle {\cal R}_c ({\bm x}) {\cal R}_c ({\bm y}) \rangle$, 
in the above equation ({\it i.e.}, only the first term).

In particular, if the power spectrum of $\chi$ is the same as that of $\phi$, the parameter $\alpha$ is
related to the local-type trispectrum parameter $\tau_{\rm NL}$ as $\tau_{\rm NL}=\alpha^2$ \cite{Byrnes:2006vq},
and Eq.~(\ref{scl}) becomes
\begin{equation}
\frac{\langle \delta^2 ({\bm x})  \delta^2 ({\bm y}) \rangle}{{\langle \delta^2 ({\bm x}) \rangle}^2}-1
\approx 4\tau_{\rm NL} \langle {\cal R}_c ({\bm x}) {\cal R}_c ({\bm y}) \rangle+{\cal O}(\alpha^4). 
\end{equation}
Thus, the correlation of the local variance is proportional to $\tau_{\rm NL}$ and the correlation function
of the curvature perturbation.
Notice that the bispectrum parameter is $f_{\rm NL}=0$ in the present case,
and it is actually the trispectrum (not the bispectrum) that determines the clustering of PBHs over the super-Hubble distance.

On the other hand, if $f_{\rm NL}\neq 0$, 
$\tau_{\rm NL}$ is also non-zero with the lower bound 
$\tau_{\rm NL} \geq \frac{36}{25} f_{\rm NL}^2$ \cite{Suyama:2007bg}. 
Thus PBHs are necessarily clustered on super-Hubble scales in this case, 
which is consistent with Refs.~\cite{Tada:2015noa, Young:2015kda} which showed 
that the clustering is characterized by $f_{\rm NL}$.

In the following discussion, we actually employ the 
local-type ansatz for the non-Gaussianities of primordial curvature perturbations.
Based on our result in Eq. (\ref{eq:PBHcorr}), 
we explicitly show that the PBH two-point correlation function is proportional to $\tau_{\rm NL}$ and the two-point correlation function of the primordial curvature perturbations.

\subsection{Primordial local-type non-Gaussianity}

The local-type primordial non-Gaussianity in Fourier space\footnote{We use the Fourier transform expression given by
\begin{equation}
f ({\bm x}) = \int \frac{d^3 k}{(2 \pi)^3} F ({\bm k}) e^{i {\bm k}\cdot {\bm x}}.    
\end{equation}}
has conventionally been characterized by introducing three constant parameters, so-called
non-linearity parameters, $f_{\rm NL},g_{\rm NL},$ and $\tau_{\rm NL}$, which respectively represent the amplitudes of the bispectrum and trispectrum
of the primordial curvature perturbations as \cite{Byrnes:2006vq}
\begin{eqnarray}
&&\langle {\mathcal R}_c ({\bm k}_1) {\mathcal R}_c ({\bm k}_2) {\mathcal R}_c ({\bm k}_3)\rangle := (2 \pi)^3 \delta^{(3)}({\bm k}_1 + {\bm k}_2 + {\bm k}_3)\, \frac{6}{5} f_{\rm NL} \left[ P_{{\mathcal R}_c}(k_1) P_{{\mathcal R}_c}(k_2) + 2~{\rm perms.} \right]~, \cr\cr
&&\langle {\mathcal R}_c ({\bm k}_1) {\mathcal R}_c ({\bm k}_2) {\mathcal R}_c ({\bm k}_3){\mathcal R}_c ({\bm k}_4)\rangle :=
(2 \pi)^3 \delta^{(3)}({\bm k}_1 + {\bm k}_2 + {\bm k}_3 + {\bm k}_4) \cr\cr
&& \qquad \qquad \qquad \qquad \qquad \qquad \quad \times \left\{
\frac{54}{25}g_{\rm NL} \left[ P_{{\mathcal R}_c}(k_1) P_{{\mathcal R}_c}(k_2) P_{{\mathcal R}_c}(k_3)  + 3~{\rm perms.} \right] \right. \cr\cr
&& \qquad \qquad \qquad \qquad \qquad \qquad \qquad 
\left. + \tau_{\rm NL} \left[  P_{{\mathcal R}_c}(k_1) P_{{\mathcal R}_c}(k_2) P_{{\mathcal R}_c}(|{\bm k}_1 + {\bm k}_3|)  + 11~{\rm perms.} \right] \right\} ~,
\label{eq:localnG}
\end{eqnarray}
where $P_{{\mathcal R}_c}(k)$ is the power spectrum of the primordial curvature perturbations given as
\begin{equation}
\langle{\mathcal R}_c ({\bm k}) {\mathcal R}_c ({\bm k}') \rangle
:= (2 \pi)^3 \delta^{(3)}({\bm k} + {\bm k}') P_{{\mathcal R}_c}(k)~.
\end{equation}

\subsection{PBH power spectrum with non-Gaussian primordial fluctuations}

In the previous section we discussed the two-point correlation function of the spatial distribution of PBHs.
The two-point correlation function
can be expressed by the PBH power spectrum as
\begin{eqnarray}
\xi_{\rm PBH}({\bm x}_1, {\bm x}_2) := \langle \delta_{\rm PBH}({\bm x}_1) \delta_{\rm PBH} ({\bm x}_2) \rangle
= \int \frac{d^3k}{(2 \pi)^3} P_{\rm PBH}(k)\, e^{i {\bm k}\cdot ({\bm x}_1 - {\bm x}_2)}~,
\end{eqnarray}
where $\delta_{\rm PBH}({\bm x})$ is the number density field of PBHs, and $P_{\rm PBH} (k)$ is the PBH power spectrum:
\begin{equation}
\langle\delta_{\rm PBH} ({\bm k}) \delta_{\rm PBH} ({\bm k}') \rangle
:= (2 \pi)^3 \delta^{(3)}({\bm k} + {\bm k}') P_{{\rm PBH}}(k)~,
\end{equation}
with $\delta_{\rm PBH}({\bm k})$ being a Fourier transform of
the PBH number density field. 
Assuming statistical isotropy, the PBH power spectrum can be inversely given by
\begin{equation}
P_{\rm PBH}(k) = \int d^3 r \, \xi_{\rm PBH} ({\bm r})
\,e^{-i{\bm k}\cdot {\bm r}}~,
\end{equation}
where ${\bm r} := {\bm x}_1 - {\bm x}_2$.
Substituting Eq.~(\ref{eq:PBHcorr}) into this equation
and noting that we apply the comoving density fluctuations, $\delta ({\bm x})$,
to $\theta_{\rm local} ({\bm x})$ used in the previous section,
we obtain
\begin{eqnarray}
P_{\rm PBH}(k) &\simeq& \frac{\nu^2}{\sigma^2_R}W_R(k)^2 P_{\delta} (k) \cr\cr
&& + \frac{\nu^3}{\sigma_R^3}\int \frac{d^3 p}{(2 \pi)^3} W_R(p) W_R(|{\bm k} - {\bm p}|) W_R(k)
B_\delta ({\bm p},-{\bm k},{\bm k} - {\bm p}) \cr\cr
&& + \frac{1}{3}\frac{\nu^4}{\sigma_R^4}\int \frac{d^3 p_1 d^3 p_2}{(2 \pi)^6}\cr\cr
&& \qquad \times W_R(p_1)W_R(p_2) W_R(|{\bm k} - {\bm p}_1 - {\bm p}_2|) W_R(k)
T_\delta ({\bm p}_1,{\bm p}_2, -{\bm k}, {\bm k} - {\bm p}_1 - {\bm p}_2) \cr\cr
&&
+ \frac{1}{4}\frac{\nu^4}{\sigma_R^4}
\int \frac{d^3 p_1 d^3 p_2}{(2 \pi)^6} \cr\cr
&& \qquad  \times
W_R(p_1)W_R(p_2) W_R(|{\bm k} - {\bm p}_1|)W_R(|{\bm p}_2 + {\bm k}|)
T_\delta ({\bm p}_1,{\bm p}_2, {\bm k}-{\bm p}_1, -{\bm k} - {\bm p}_2)~, \nonumber\\
\label{eq:Ppbh1}
\end{eqnarray}
where 
$P_\delta,B_\delta,~$and $T_\delta$ are respectively
power, bi-, and trispectra of the comoving density fluctuations,
$W_R(k)$ is a window function smoothed with the comoving scale
$R = (aH)^{-1}$ at the PBH formation in Fourier space.

For primordial local-type non-Gaussianity
as defined by Eq. (\ref{eq:localnG}),
$P_\delta,B_\delta,~$and $T_\delta$ are respectively
given in terms of the non-linearity parameters and the power spectrum of the primordial curvature perturbations as
\begin{eqnarray}
P_\delta (k) &=&  \left(\frac{4}{9}\right)^2 (kR)^4  P_{{\mathcal R}_c}(k), \cr\cr
B_\delta (k_1, k_2, k_3) &=& \left(\frac{4}{9}\right)^3 (k_1R)^2 (k_2R)^2 (k_3R)^2  \frac{6}{5}f_{\rm NL}
 \left[ P_{{\mathcal R}_c}(k_1) P_{{\mathcal R}_c}(k_2) + 2~{\rm perms.} \right], \cr\cr
T_\delta ({\bm k}_1, {\bm k}_2, {\bm k}_3, {\bm k}_4) &=& 
\left(\frac{4}{9}\right)^4 (k_1R)^2 (k_2R)^2 (k_3R)^2(k_4R)^2 \cr\cr
&& \times \left\{
\frac{54}{25}g_{\rm NL} \left[ P_{{\mathcal R}_c}(k_1) P_{{\mathcal R}_c}(k_2) P_{{\mathcal R}_c}(k_3)  + 3~{\rm perms.} \right] \right. \cr\cr
&&\qquad  \left. + \tau_{\rm NL} \left[  P_{{\mathcal R}_c}(k_1) P_{{\mathcal R}_c}(k_2) P_{{\mathcal R}_c}(|{\bm k}_1 + {\bm k}_3|)  + 11~{\rm perms.} \right] \right\}~.
\label{eq:Deltaspec}
\end{eqnarray}
Substituting Eq. (\ref{eq:Deltaspec}) into Eq. (\ref{eq:Ppbh1}), we have
\begin{eqnarray}
P_{\rm PBH}(k) &\simeq& \left(\frac{4\nu}{9\sigma_R}\right)^2 W_{\rm local}(k)^2 P_{{\mathcal R}_c} (k) \cr\cr
&&
+ \frac{6}{5}f_{\rm NL} \left(\frac{4\nu}{9\sigma_R} \right)^3 
W_{\rm local}(k) \cr\cr
&& \times
\int \frac{d^3 p}{(2 \pi)^3} W_{\rm local}(p) W_{\rm local}(|{\bm k} - {\bm p}|) \cr\cr
&& \quad \times 
\left[ 2 P_{{\cal R}_c}(p) P_{{\cal R}_c}(k) + P_{{\cal R}_c}(p) P_{{\cal R}_c}(|{\bm k} - {\bm p}|) \right] \cr\cr
&& + \frac{18}{25}g_{\rm NL}\left(\frac{4\nu}{9\sigma_R} \right)^4 W_{\rm local}(k) \cr\cr
&& \times
\int \frac{d^3 p_1 d^3 p_2}{(2 \pi)^6}W_{\rm local}(p_1)W_{\rm local}(p_2) W_{\rm local}(|{\bm k} - {\bm p}_1 - {\bm p}_2|) \cr\cr
&& \quad\quad \times \left[ 3 P_{{\mathcal R}_c}(k)P_{{\mathcal R}_c}(p_1)P_{{\mathcal R}_c}(p_2) + P_{{\mathcal R}_c}(p_1)P_{{\mathcal R}_c}(p_2)P_{{\mathcal R}_c}(|{\bm k} - {\bm p}_1 - {\bm p}_2|) \right] \cr\cr
&&
 + \frac{\tau_{\rm NL}}{3}\left(\frac{4\nu}{9\sigma_R} \right)^4 W_{\rm local}(k) \cr\cr
&& \times
\int \frac{d^3 p_1 d^3 p_2}{(2 \pi)^6}W_{\rm local}(p_1)W_{\rm local}(p_2) W_{\rm local}(|{\bm k} - {\bm p}_1 - {\bm p}_2|) \cr\cr
&& \quad\times \left[ 6 P_{{\mathcal R}_c}(k)P_{{\mathcal R}_c}(p_1)P_{{\mathcal R}_c}(|{\bm p}_1 + {\bm p}_2|) + 6 P_{{\mathcal R}_c}(p_1)P_{{\mathcal R}_c}(p_2)P_{{\mathcal R}_c}(|{\bm k} - {\bm p}_1 |) \right] \cr\cr
&&
+ \frac{54}{25}g_{\rm NL}\left(\frac{4\nu}{9\sigma_R}\right)^4
 \cr\cr
&& \times \int \frac{d^3 p_1 d^3 p_2}{(2 \pi)^6}W_{\rm local}(p_1)W_{\rm local}(p_2) W_{\rm local}(|{\bm k} + {\bm p}_1|)W_{\rm local}(|{\bm p}_2 - {\bm k}|) \cr\cr 
&& \qquad \qquad \qquad \times  P_{{\mathcal R}_c}(p_1) P_{{\mathcal R}_c}(p_2) P_{{\mathcal R}_c}(|{\bm k}+{\bm p}_1|) \cr\cr
&&
+ \frac{\tau_{\rm NL}}{4}\left(\frac{4\nu}{9\sigma_R}\right)^4 \cr\cr
&& \times 
\int \frac{d^3 p_1 d^3 p_2}{(2 \pi)^6}W_{\rm local}(p_1)W_{\rm local}(p_2) W_{\rm local}(|{\bm k} + {\bm p}_1|)W_{\rm local}(|{\bm p}_2 - {\bm k}|) 
 \cr\cr
&& \times \left[ 4 P_{{\cal R}_c}(p_1)P_{{\cal R}_c}(p_2) P_{{\cal R}_c}(k) \right.\cr\cr
&& \quad \left.
+ 4 P_{{\cal R}_c}(p_1) \left( P_{{\cal R}_c}(p_2) P_{{\cal R}_c}(|{\bm k} + {\bm p}_1 - {\bm p}_2|) + P_{{\cal R}_c} (|{\bm k} + {\bm p}_1|)P_{{\cal R}_c}(|{\bm p}_1 + {\bm p}_2|) \right) \right]~, \nonumber\\
\label{eq:Ppbh2}
\end{eqnarray}
where $W_{\rm local}(k) := (kR)^2 \,W_R(k)$.
As can be seen in the above equation, the contributions from non-zero $g_{\rm NL}$ and $\tau_{\rm NL}$ are of the order of $P_{{\mathcal R}_c}^3$.
Note that this equation is derived based on the approximate expression (\ref{eq:PBHcorr}).
With respect to the order of $P_{{\mathcal R}_c}$, 
$(\xi^{(2)})^2$, $(\xi^{(2)})^3$,
and $\xi^{(2)} \xi^{(3)}$ terms should also be of the  
order of $P_{{\mathcal R}_c}^3$
and hence they should be 
taken into account in the expression in Eq. (\ref{eq:PBHcorr}) (or Eqs.~(\ref{eq:exP1_a}) and (\ref{eq:exP2_a})).
However, one can show that these quadratic and cubic terms are
included in the $k$-independent contribution as in the following expression. Thus, for the super-Hubble correlation of PBHs,
these terms can be neglected.

For the $kR \ll 1$ limit, which we focus on in this paper, 
noting that we can take $W_{\rm local}(k) \to 0$ and $|{\bm k}+ {\bm p}| \to p$,
the above expression becomes 
\begin{eqnarray}
P_{\rm PBH}(k) &\simeq & \tau_{\rm NL} \left(\frac{4\nu}{9\sigma_R}\right)^4
\left[\int \frac{d^3p}{(2\pi)^3} W_{\rm local}(p)^2 P_{{\cal R}_c}(p) \right]^2 P_{{\cal R}_c}(k) + ~(k~{\rm -}~{\rm independent}~{\rm terms}) \cr\cr
&=& \tau_{\rm NL} \nu^4 P_{{\cal R}_c}(k) + ~(k~{\rm -}~{\rm independent}~{\rm terms})~.
\label{eq:powerPBH}
\end{eqnarray}
Thus, in the case where the primordial curvature perturbations have local-type
non-Gaussianity parameterized by $\tau_{\rm NL}$,
the PBH power spectrum does not decay even on super-Hubble scales and
is proportional to the power spectrum of the primordial curvature perturbations.
Inversely, the two-point correlation function of the PBHs is obtained as
\begin{eqnarray}
\xi^{(2)}_{\rm PBH}({\bm r}) 
= \int \frac{d^3 k}{(2 \pi)^3} P_{\rm PBH}(k) e^{i{\bm k}\cdot{\bm r}} 
= \tau_{\rm NL} \nu^4 \xi^{(2)}_{{\mathcal R}_c} ({\bm r})
+C\,\delta^{(3)}({\bm r}),
\end{eqnarray}
where a constant, $C$, corresponds to the $k$-independent terms in Eq. (\ref{eq:powerPBH}). 
A typical value of the enhancement factor is
$\tau_{\rm NL} \nu^4 \sim O(10^6) \times (\tau_{\rm NL}/10^3)$.
For super-Hubble scales, only the first term on the right hand side
is relevant, and
thus, if $\tau_{\rm NL}$-type non-Gaussianity exists, it would give a large effect
on the clustering behavior of PBHs even on super-Hubble scales ($r \gg R$) at formation.

As discussed in Refs.~\cite{Tada:2015noa,Young:2015kda},
if PBHs contribute to the dark matter component of the Universe,
their spatial distribution behaves as dark matter isocurvature perturbations. Recent cosmic microwave background (CMB) observations give a tight constraint on the fraction of  dark matter isocurvature perturbations~\cite{Akrami:2018odb}.
By introducing a parameter representing the mass fraction of PBHs in the total dark matter, $f_{\rm PBH}$, the observational constraint roughly means $f_{\rm PBH}^{2} \tau_{\rm NL} \nu^4 \lesssim O(10^{-2})$.
If PBHs comprise a dominant component of dark matter,
adopting $\nu=10$ as an approximate value necessary for PBH formation gives
an upper limit on $\tau_{\rm NL}$ of $\tau_{\rm NL} \lesssim 10^{-6}$.

Note that this constraint for $\tau_{\rm NL}$ is obtained by assuming that the simple expression for the primordial trispectrum
given by Eq. (\ref{eq:localnG}) is valid for CMB scales which are super-Hubble ones at the PBH formation.
Although the slow-roll condition would be violated when PBH formation scale exits the horizon, in general,
the CMB scales are considered to exit the horizon in the slow-roll phase.
As discussed in e.g. Ref.~\cite{Cabass:2016cgp},
it might be difficult to realize simple expressions for the primordial non-Gaussianity given by Eq. (\ref{eq:localnG}) in single-field slow-roll inflation.
Thus, the above constraint for $\tau_{\rm NL}$ cannot be simply applied in single-field inflation
and it does not tightly constrain the PBH formation scenario in such a single-field class.

\section{Conclusion}
\label{sec:con}

We have investigated the clustering behavior of PBHs at  formation during the radiation-dominated era.
Since during the radiation-dominated era PBHs would be formed from the direct collapse of the overdense region
with Hubble scales, for PBH clustering, super-Hubble fluctuations might be important.
We formulated the two-point correlation function of PBHs by making use of a functional integration approach which takes into account the non-Gaussian property of the primordial fluctuations.
Our result shows that PBHs are never clustered at formation 
as long as the primordial fluctuations obey Gaussian statistics and the super-Hubble two-point correlation function
is induced by the connected part of the four-point correlation function of the primordial fluctuations.
In order to evaluate the super-horizon two-point correlation function of PBHs quantitatively, 
we considered non-Gaussian primordial perturbations of the local type up to
the four-point function(trispectrum) parametrized by two constant parameters, $g_{\rm NL}$ and $\tau_{\rm NL}$.
We found that the $\tau_{\rm NL}$-type non-Gaussianity determines the super-Hubble two-point correlation function. 
Thus, to estimate the clustering behavior of PBHs
we should carefully investigate the non-Gaussian property of the primordial fluctuations which strongly depends on the generation mechanism, such as the inflation model.

\section*{Acknowledgments}

 We would like to thank Misao Sasaki, Tomohiro Harada and Chul-Moon Yoo for useful discussions.
 SY would also like to thank Jinn-Ouk Gong and Takahiko Matsubara for useful discussions about the functional integration approach formulation. 
 SY is supported by MEXT KAKENHI Grant Numbers 15H05888
and 18H04356.
TS is supported by JSPS Grant-in-Aid for Young Scientists (B) No.15K17632, and by the MEXT Grant-in-Aid for Scientific Research on Innovative Areas Nos.15H05888, 17H06359, 18H04338, and 19K03864.


\bibliographystyle{unsrt}
\bibliography{main}
%

\end{document}